%% file: paper.resid_symm.tex
\documentstyle[amssymb]{europhys}
 
\newcommand{\bea}{\begin{eqnarray}}
\newcommand{\eea}{\end{eqnarray}}

\newcommand{\ket}[1]{| #1 \rangle}
\newcommand{\braket}[2]{\langle {#1} |{#2} \rangle}

\def\d2d2r#1{ {d^2 {#1}}\over{dr^2}}

\def\ddr{\frac{\partial}{\partial r}}
\input epsf
\input euromacr.tex

\begin{document}
\euro{}{}{}{}
\Date{}
\shorttitle{A. Krug and A. Buchleitner Residual Symmetries in the Spectrum 
of Periodically Driven Alkali Rydberg States}
\title{Residual Symmetries in the Spectrum of Periodically Driven Alkali 
Rydberg States}
\author{Andreas Krug\inst{1,2,3} and Andreas Buchleitner\inst{1,3}}
\institute{
	\inst{1} Max-Planck-Institut f\"ur Physik komplexer Systeme, 
		N\"othnitzer-Str. 38, D-01069 Dresden;
	\inst{2} Max-Planck-Institut f\"ur Quantenoptik,
Hans-Kopfermann-Str. 1, D-85748 Garching b. M\"unchen;
	\inst{3} Sektion Physik der Ludwig-Maximilians-Universit\"at 
M\"unchen, Schellingstr. 4, D-80799 M\"unchen.}

\rec{}{}
\pacs{
\Pacs{32}{80Rm}{Multiphoton ionization and excitation to highly excited states
(e.g., Rydberg states)}
\Pacs{05}{45+b}{Theory and models of chaotic systems}
\Pacs{42}{50Hz}{Strong-field excitation of optical transitions in quantum
systems; multi-photon processes; dynamic Stark shift}
	}
\maketitle
\begin{abstract}
We identify a 
fundamental structure in the spectrum of microwave driven alkali
Rydberg states, which highlights the remnants of the Coulomb symmetry in the 
presence of a non-hydrogenic core. Core-induced corrections with respect to 
the hydrogen spectrum can be accounted for by a perturbative approach.

\end{abstract}

\section{Introduction}
The excitation and subsequent ionization of Rydberg states of atomic hydrogen 
by microwave fields is one of the most prominent examples of the manifestation
of classically nonlinear dynamics in a realistic physical system \cite{pmk}. 
Given a
driving field frequency comparable to the classical Kepler frequency of the
unperturbed Rydberg electron, the electron's classical
trajectory goes chaotic for sufficiently large driving field amplitudes, 
finally leading to its ionization on a finite time scale \cite{cas}. 
Correspondingly, large
ionization rates are observed in experiments on real (i.e., quantum) Rydberg
states of atomic hydrogen, in the appropriate parameter range \cite{pmk,bay1}. 

As a matter of fact, already before the onset of classically chaotic motion,
i.e. at not too large driving field amplitudes, individual quantum eigenstates
of the atom in the field exhibit energies and ionization rates which are
determined only by the orbital parameters of the classical
trajectory they are associated with 
\cite{abu97}. Those
orbits which are the least stable under the external perturbation (i.e., which
turn chaotic for the lowest values of the driving field amplitude, such as
straight line orbits parallel to the field polarization axis for a linearly
polarized drive) induce the largest ionization rates for their associated
eigenstates. Consequently, in this near-integrable regime of classical
dynamics, it is possible to classify the eigenstates of the
atom in the field through quantum numbers associated with the orbital
parameters of unperturbed Kepler ellipses, i.e. with the angular momentum and
the  Runge-Lenz vector. An adiabatic invariant governs the slow evolution of
these parameters under external driving \cite{abu97}.

It should be noted, however, that a considerable part of experimental data
has been accumulated in experiments on Rydberg states of alkali atoms rather
than of atomic hydrogen \cite{pillet84,gallagher89,fu90,blu,arndt91,benson95}. 
A priori, a classical-quantum correspondence as
briefly sketched above for atomic hydrogen cannot be established here, due to
the absence of a well and uniquely defined classical Hamiltonian. In
particular, the atomic core destroys the symmetry characteristic for the
hydrogen atom and the Runge-Lenz vector is no more a constant of motion. 

Indeed,
experimental data systematically suggest strongly enhanced ionization rates of
nonhydrogenic (i.e., low angular momentum) alkali Rydberg states as compared
to atomic hydrogen \cite{pillet84,gallagher89,fu90,arndt91,benson95}, 
though they also 
exhibit 
qualitatively similar features,
e.g. of the dependence of the ionization yield on the principal quantum number
of the atomic state the atoms are initially prepared in 
\cite{arndt91,benson95}. 
On the other hand, a direct
comparison of available hydrogen and alkali data is somewhat questionable,
since relevant experimental parameters such as the interaction time of the
atom with the field are typically different for different
experiments. Furthermore,
a rigourous theoretical treatment of alkali atoms exposed to microwave
fields was not accomplished until now.

It is the purpose of the present letter to outline such a rigourous treatment
which allows for the first time for a {\em direct} comparison of hydrogen and
alkali ionization dynamics under {\em precisely the same} conditions, without
adjustable parameters. First results of our numerical experiments directly
address the above question of quantum-classical correspondence for
periodically driven alkali atoms. 

\section{Theory}
Let us start with the nonrelativistic 
Hamiltonian of a one-electron atom exposed to a linearly polarized microwave
field of 
(constant) amplitude $F$ and frequency $\omega$, in length gauge, employing
the dipole approximation and atomic units:
\bea
H(t)={{\bf p}^2\over 2}+V_{\rm atom}(r)+Fz \cos \omega t,\  r>0.
\label{hamil}
\eea
As this Hamiltonian is 
periodic in time, 
we can use the Floquet theorem \cite{shirley} to 
find the eigenstates (``dressed states'') of the atom in the field. 
After integration over the solid angle we have to solve 
the time-independent, radial eigenvalue equation
\begin{eqnarray}
& & \left(- {\d2d2r{}} + {\frac{\ell\left(\ell+1\right)}{r^2}}+2V_{\rm atom}
\left(r\right) -2k \omega -2 \varepsilon\right)
\ket{\Psi_{\varepsilon,\ell}^k}\nonumber \\
& & +FrA_{\ell+1}\left(\ket{\Psi_{\varepsilon,\ell+1}^{k-1}}
+\ket{\Psi_{\varepsilon,\ell+1}^{k+1}}
\right)+FrA_{\ell}\left(\ket{\Psi_{\varepsilon,\ell-1}^{k-1}}
+\ket{\Psi_{\varepsilon,\ell-1}^{k+1}}
\right)=0,
\nonumber\\
& & {\rm with}\ A_{\ell}=\sqrt{\frac{\ell^2-m^2}{4\ell^2-1}};\
\ell=0,1,2,\ldots;\ k= -\infty, \ldots, +\infty .  \label{dgl}
\end{eqnarray}
The additional quantum number $k$ counts the number of photons that are
exchanged between the atom and the field, and $\varepsilon$ 
denotes the quasi-energy of the dressed state 
\begin{equation}
\ket{\Psi_{\varepsilon}}=\sum_k\exp(-ik\omega t)\ket{\Psi_{\varepsilon}^k}=
\sum_{k,\ell} \exp(-ik\omega t) 
Y_{\ell,m}(\theta,\phi)\ket{\Psi_{\varepsilon,\ell}^k}/r,
\end{equation}
with $Y_{\ell,m}(\theta,\phi)$ the spherical harmonics.
$m$ denotes the angular momentum projection on the field polarization axis and
remains 
a good quantum number, due to the rotational symmetry of our problem around
the field axis. For all numerical results presented hereafter, its value was
fixed to $m=0$.
As immediately obvious from the nondiagonal part of 
eq.~(\ref{dgl}), 
the interaction with the linearly polarised microwave field 
conserves the generalised parity $\Pi=(-1)^{k+\ell}$.
This just expresses the angular momentum transfer associated with the 
absorption (emission) of a photon.

As a unique one-particle 
potential $V_{\rm atom}(r)$ for alkali atoms is unknown, we use a
variant \cite{halley93} of 
R-matrix 
theory to describe the interaction of the outer electron with the atomic core.
Configuration space is divided in two regions: 
In the internal region, 
$0<r\leq a$, the external field is negligible compared to the 
field
created by the atomic core, and the details of the interaction are unknown.
With the help of quantum defect theory~\cite{seaton83}, the solution 
of eq.~(\ref{dgl}) at $r=a$ 
can be written as a linear combination of regular and irregular 
Coulomb-functions $s_{\ell,E}(r)$ and $c_{\ell,E}(r)$, 
\bea
F_{\ell,E}(r)=\cos(\pi \delta _\ell)s_{\ell,E}(r)+\sin(\pi 
\delta _\ell)c_{\ell,E}(r),\ r=a,
\label{innerfct}
\eea
where the $\delta_{\ell}$ are the quantum defects~\cite{seaton83} known from 
spectroscopic experimental data \cite{lorenzen}.
In the outer region, 
$ r > a$,
the difference between the actual atomic potential $V_{\rm atom}(r)$ and the 
Coulomb potential
$-1/r$ can be neglected. However, the operator 
$d^2 /dr^2$ is no more hermitian in the reduced range $a < r < \infty$. 
To overcome 
this problem, a surface term $\delta (r-a)(\ddr +C_\ell)$ is 
added~\cite{halley93,bloch} to the diagonal part of (\ref{dgl}). The 
matching condition between inner and outer region 
at $r=a$ is incorporated in the constant  
$C_\ell$ by defining
\bea
C_\ell =(F_{\ell,\varepsilon +k\omega}(r))^{(-1)}{\ddr F_{\ell,\varepsilon
+k\omega}(r)}.\label{surf}
\eea
Note that the function $F_{\ell,E}(r)$ in eq.~(\ref{innerfct}) has to be 
evaluated at the energy $\varepsilon +k \omega$ in (\ref{surf}), i.e. at
different energies for different photon 
indices $k$. This generalizes the
approach outlined in \cite{halley93} to periodically driven systems.

Finally, due to the continuum coupling induced by 
the external 
field, all atomic 
bound states turn into resonances with finite ionization rates
$\Gamma_{\epsilon}$. 
In order to
extract the latter together with the
energies $\epsilon$ 
of the atom in 
the field, we use the method of complex scaling~\cite{abu95,balslev}. 
After this nonunitary transformation the Floquet Hamiltonian 
amended by the 
core induced surface term (\ref{surf}) 
is represented by a complex symmetric matrix, with complex eigenvalues 
$\varepsilon -i \Gamma_{\varepsilon}/2$.
These are obtained by diagonalization
of the complex eigenvalue problem in a real Sturmian basis, using an efficient
implementation of the Lanczos algorithm. Together with the associated
eigenvectors they provide a complete description of our problem \cite{abu95}.

\section{Results}
\begin{figure}[b]
  \begin{minipage}[t]{148mm}
    \begin{minipage}[t]{85mm}
      \epsfysize=64mm {\epsfbox{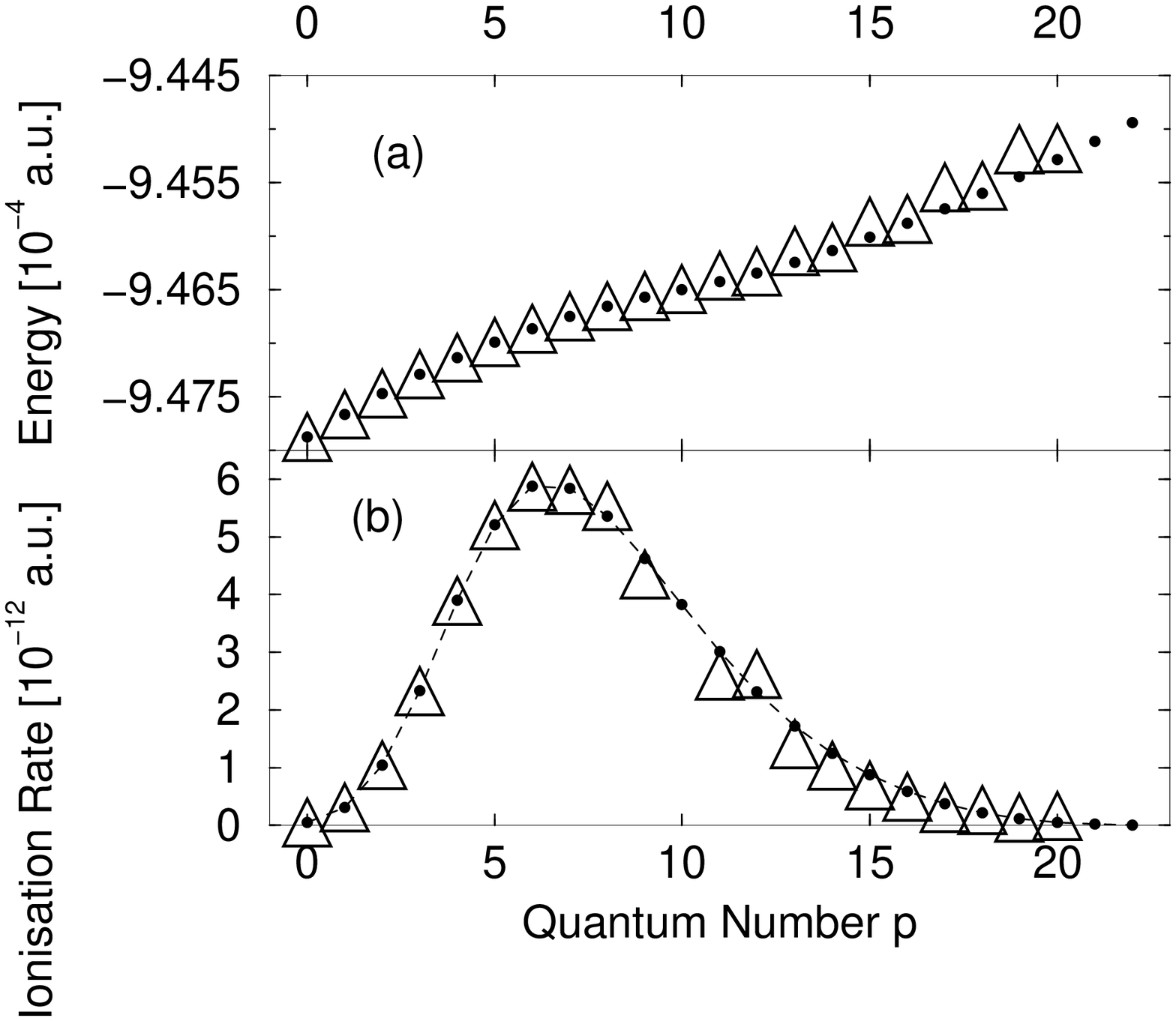}}
    \end{minipage}
    \begin{minipage}[t]{85mm}
      \epsfysize=62mm {\epsfbox{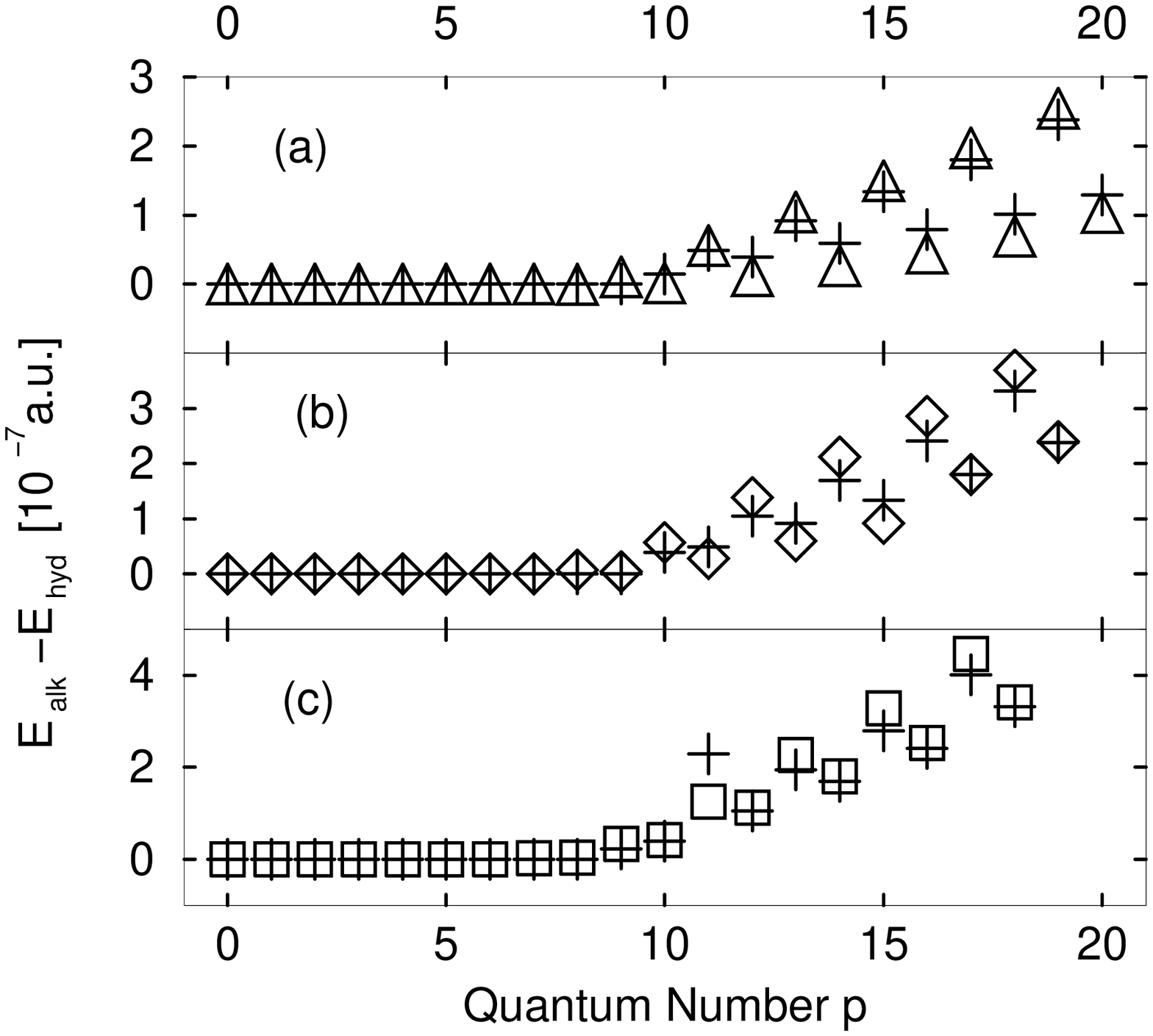}}
    \end{minipage}
  \end{minipage}
\caption{Energies (a) and ionisation rates (b) of 
Rydberg states of lithium (triangles) and of atomic hydrogen (dots) exposed to
a linearly polarized microwave field of frequency $\omega = 1.07171794 \times
10^{-4} {\rm a.u.}$ and amplitude $F=1.072 \times 10^{-7} {\rm a.u.}$, for
principal quantum number $n_0=23$ and angular momentum projection $m=0$ on the
field polarization axis. The lithium spectrum lacks two of the 23 substates of
the manifold, due to the quantum defects $\delta_{\ell=0}=0.399468$ and
$\delta_{\ell=1}=0.047263$ of the $\ell=0$ and $\ell=1$ states,
respectively. The quantum defects $\delta_{\ell=2}=0.002129$ and
$\delta_{\ell=3}=-0.000077$ are negligible compared to the field induced
splitting of the $n_0=23$ manifold (field-free energy $E_{23}\simeq
-9.452\times 10^{-4}\ \rm a.u.$). Both spectra almost coincide (in energy
and ionisation rate) even for larger values ($p\geq 10$) of the (semiclassical
\protect\cite{abu97}) quantum number $p$, despite the fact that the
localization properties of the associated eigenstates (close to the plane
defined by the field polarization axis) originate in the dynamical symmetry
of the $-1/r$ Coulomb potential \protect\cite{dd87}. The latter is {\em
destroyed} by the presence of a nonhydrogenic core in alkali atoms. The
ionization rate of the $p=10$ state of lithium is locally enhanced by
approx. one order of magnitude with respect to the corresponding hydrogen
eigenstate, due to a six-photon resonance with the $|n=41, \ell=0\rangle$ 
state.}
\label{fig1}
\caption{Shifts $E_{\rm alk}-E_{\rm hyd}$ of the energies $E_{\rm alk}$ of 
lithium (a, triangles), sodium (b, diamonds), and rubidium (c, squares) 
as compared to those, $E_{\rm hyd}$, of 
the $n_0=23$ manifold of atomic hydrogen 
in a linearly
polarized microwave field, with the same parameters as in
fig.~\protect\ref{fig1}. Quantum defects employed for the sodium results:
$\delta_{\ell=0}=1.347964$, $\delta_{\ell=1}=0.85538$,
$\delta_{\ell=2}=0.015543$, $\delta_{\ell=3}=0.001453$, and for rubidium: 
$\delta_{\ell=0}=3.1311$, $\delta_{\ell=1}=2.6415$, $\delta_{\ell=2}=1.3472$,
$\delta_{\ell=3}=0.016312$ \protect\cite{lorenzen}. Consequently, three
respectively four energy levels are missing in (b) and (c). The nonvanishing 
shifts 
for large $p\geq 9$ values can be
accounted for by projecting out the low $\ell$ components (i.e. the ones with
core induced energy shifts large with respect to the field induced splitting
of the $n_0=23$ manifold) of the $n_0$-manifold, as indicated by the
crosses, see eqs.~(8) and (9). The agreement between this perturbative
approach and the exact quantum results is always better than the average
level spacing of the hydrogen manifold (dots in fig.~1), except for the 
relatively 
large 
discrepancy at $p=11$, in (c). The latter is due to a
multiphoton resonance
between the 
alkali
eigenstate 
and a nonhydrogenic (low $\ell$) state.}
\label{fig2}
\end{figure}

The described theoretical/numerical 
apparatus is now applied to alkali atoms in a microwave field. Since we want 
to identify the core induced effects in the alkali problem as compared to the
hydrogen spectrum, we use parameter values which have been employed
in earlier work on microwave driven Rydberg states \cite{abu97,abu95} of
hydrogen. To keep the comparison as transparent as possible, we focus on a
microwave frequency $\omega =1.07171794\times 10^{-4} {\rm \ a.u.}$ which is
nonresonant with the hydrogen level spacing in the vicinity of the atomic 
initial
state with principal quantum number $n_0=23$. The field amplitude is fixed to 
$F=1.072\times 10^{-7} {\rm \ a.u.}$, slightly below the onset of appreciable
(chaos-induced \cite{cas}) ionization of atomic hydrogen \cite{abu97}. 
This
choice of parameters defines a near-integrable phase space structure for the
classical dynamics of driven hydrogen, with an unambiguous signature in the
associated quantum energies emerging from the $n_0=23$ manifold. The black
dots in fig.~\ref{fig1} illustrate the situation: The driving field lifts
the angular momentum degeneracy of the substates of the manifold, which
reorganize according to their localization properties in classical phase space
\cite{abu97}. Those states with maximum angular momentum and 
spherical symmetry  
experience the
strongest field induced (``ac-'') shift in energy, whereas those with maximum
radial component of the Runge-Lenz vector and ``$\lambda$-symmetry''
\cite{abu97,dd87,ddt} remain essentially unaffected by the external
perturbation. Since the low angular momentum states are strongly mixed by the
field (to build states with $\lambda$-symmetry \cite{dd87,ddt}), a new
(semiclassical) quantum number $p$ \cite{abu97} 
is used to label the $n_0$ substates of the
manifold in the field. $p$ is an integer ranging from $0$ to $n_0-1$, and 
simply counts the number of quanta enclosed by a semiclassical contour integral
along the equipotential curves of the adiabatic Hamiltonian which generates the
slow evolution of angular momentum and Runge-Lenz vector of the classical 
Kepler ellipse under external driving \cite{abu97}. The
associated eigenstates exhibit spherical symmetry for $p=0\ldots 9$, and 
$\lambda$-symmetry for $p=10\ldots 22$, respectively \cite{abu97}.
Note that low and high $p$-values correspond to
negligible 
ionization rates of the atom in the field, due to the {\em classical}
stability of the associated trajectories under external driving
\cite{abu97}. Actually, the $\lambda$-states with large $p$, which quantize a
classical straight line orbit perpendicular to the field polarization axis,
with maximum modulus of the Runge-Lenz vector, display the smallest ionization
rates \cite{abu97}.

In the presence of a non-hydrogenic core, the Runge-Lenz vector is no more a
conserved quantity and the $\lambda$-symmetry defining associated eigenstates
of the field free atom \cite{dd87} 
is destroyed. Therefore, no symmetry argument is
available to predict a similar (semiclassical) organization of the alkali
energy levels under external driving, alike the one observed for atomic
hydrogen \cite{abu97}.

Nonwithstanding, our results for lithium Rydberg states exposed to precisely
the same external perturbation as for the hydrogen results clearly show that
the symmetry properties of the driven Coulomb problem prevail even in the
presence of the core. As evident from the open triangles in 
fig.~\ref{fig1} (a), the hydrogenic 
part of the lithium manifold exhibits globally
the same (semiclassical) structure as the hydrogen levels. For low values of
$p$ ($\simeq 0\ldots 9$) this is not surprising as the associated classical
trajectories (large angular momenta) do not probe the atomic
core \cite{abu97}. 
However, for large $p$-values ($\simeq 10\ldots 20$), the classical
solution of the Coulomb problem does impinge on the nucleus and will certainly
suffer scattering off 
the nonhydrogenic core. Yet, in the presence of the field,
this scattering obviously 
mixes 
states of $\lambda$ type only and
does not affect the overall separation of the spectrum in spherical and
$\lambda$ states, as a remnant of the classical phase space structure of the
driven Coulomb dynamics. Neither does the presence of the core appreciably
affect the ionization rates of the dressed states, as obvious from 
fig.~\ref{fig1} (b). Only at $p=10$
is there a local enhancement of
the width (by approx. one order of magnitude), 
due to the near resonant coupling of the state to the nonhydrogenic
eigenstate 
originating from $|n=41, \ell=0\rangle$, 
via a
six-photon transition (similarly, a very weak mutliphoton coupling slightly
enhances the width of the $p=12$ state). 
In the near integrable regime of the classical Coulomb
dynamics we are considering here it is precisely this kind of multiphoton
resonances between nonhydrogenic (low $\ell$, such that $\delta_{\ell}\ne 0$) 
states and hydrogenic manifolds
which provides a channel for enhanced ionization as compared to atomic
hydrogen. Note that without such a near resonant coupling, the non-hydrogenic
states of a given manifold tend to be {\em more} stable than the hydrogenic
ones, as they are highly isolated in the spectrum. As an example, for the 
same field parameters, the lithium 
$n_0=23$
$\ell=0$ ($\delta_{\ell=0}=0.399468$) 
and $\ell=1$ ($\delta_{\ell=1}=0.047263$) \cite{lorenzen} 
states exhibit ionization rates
$\Gamma_{\varepsilon}\sim 10^{-15}\ \rm a.u.$ as small as the most
stable substates of the hydrogenic manifold of fig.~\ref{fig1}. A detailed
analysis of enhanced ionization via core-induced multiphoton resonances will
be provided elsewhere.

Closer inspection of fig.~\ref{fig1} (a) shows additional structure in
the alkali spectrum, on top of the globally hydrogen-like structure: for large
values of $p$ ($\geq 11$), the alkali levels are shifted with respect to the
hydrogenic energies. These shifts can be recovered by diagonalization of the
hydrogen problem within the restricted subspace spanned by the hydrogenic
levels of the alkali Rydberg manifold \cite{ddt,fabre,braun85,penent}. 
In other words, the shifted energies
are the solutions of the eigenvalue equation 
\bea
PH_{\rm hyd}P\ket{\Phi^{k_0}}=(E+k_0\omega)\ket{\Phi^{k_0}}, \label{proj}
\eea
where $H_{\rm hyd}$ is  
obtained from 
from (1) setting 
$V_{\rm atom}(r)=-1/r,\
r\in]0,\infty[$, and 
$P$ the projector onto the hydrogenic subspace of the alkali manifold
labeled by the principal quantum number $n_0$ and the photon number $k_0$. 
Such a procedure is legitimate as long as the states
emerging from the nonhydrogenic part of the alkali manifold have vanishing
overlap with the complete hydrogen manifold emanating from $(n_0,k_0)$. This
condition is fulfilled for the driving field strength considered here.

Solving (\ref{proj}) for $E$ is tantamount to finding the roots of 
\bea
\det(Q{\frac{1}{H_{\rm hyd}-(E+k_0\omega)}}Q)=0,
\label{determ}
\eea
with $Q=1-P$ the projector onto the orthogonal complement of the hydrogenic
subspace for given $(n_0,k_0)$. Without loss of generality we choose $k_0=0$ 
hereafter. Consequently, for one single non-vanishing quantum defect
$\delta_{\ell_0}$, (\ref{determ}) becomes 
\bea
\sum_{\varepsilon}\frac{|{{\braket{n_0,\ell_0}{\Psi^{k_0=0}_\varepsilon}}}|^2}
{\varepsilon -E}=0,
\label{det1state}
\eea
where $|n_0,
\ell_0\rangle$ spans the orthogonal 
complement of the hydrogenic subspace of the alkali
atom within the $(n_0,k_0=0)$ manifold. Note that (\ref{determ}) or 
(\ref{det1state})
have to be evaluated for different values of the generalized parity $\Pi$,
and that we have to solve (\ref{det1state}) separately for $\ell_0=0$ and
$\ell_0=1$, in order to recover the level shifts observed for lithium 
in fig.~\ref{fig1} (the $\ell_0=2$ and $\ell_0=3$ states of lithium remain
within the range of $P$, due to their negligible quantum defects 
$\delta_{\ell=2}=0.002129$ and $\delta_{\ell=2}=-0.000077$ \cite{lorenzen}, 
at the given
field strength). Fig.~\ref{fig2} (a) shows the result of the projection
method, compared to the exact numerical result -- the agreement is very good.
Since the low $p$ states essentially exhibit spherical symmetry with large
angular momentum projection, their overlap with 
$|n_0, \ell_0=0 (\ell_0=1)\rangle$
vanishes and their energies remain unshifted as compared to the hydrogen
results. 

The scenario which we described for lithium also applies for the heavier
alkali elements, as illustrated in figs.~\ref{fig2} (b) and (c). Here we plot
the shifts of the 
exact energies of sodium and rubidium with respect to the hydrogen levels,
as they emerge from the $n_0=23$
manifold, for precisely the same field parameters as used for the lithium
results. Since for these elements also the $\ell_1=2$ (sodium) and the
$\ell_1=3$ (rubidium) states are separated from the hydrogenic manifold due to
their large quantum defects, the range of $Q$ in (\ref{determ}) is
two-dimensional and the evaluation of the determinant yields the
expression
\begin{equation}
\sum_{\varepsilon}\frac{|{\braket{n_0,\ell_0}{\Psi^{k_0=0}_\varepsilon}}|^2}
{\varepsilon -E}\sum_{\varepsilon}\frac{|{\braket{n_0,\ell_1}
{\Psi^{k_0=0}_\varepsilon}}|^2}{\varepsilon -E}-
\left[\sum_{\varepsilon}\frac{{\braket{n_0,\ell_0}{\Psi^{k_0=0}_\varepsilon}}
{\braket{\Psi^{k_0=0}_\varepsilon}{n_0,\ell_1}}}{\varepsilon -E}
\right]^2
=0.
\label{resolvtwostates}
\end{equation}
Again, the solution of (\ref{resolvtwostates}) 
gives very good agreement with the numerical
result. In addition, we note that the larger the dimension of the range of
$Q$, the smaller the values of $p$ for which the alkali levels are shifted as
compared to the hydrogen energies. This is a consequence of the dominance of
small $\ell$ components in large $p$ states and of large $\ell$ components in
small $p$ states, since the heavier the element the larger the $\ell$ values
affected by non-negligible quantum defects. 

\section{Summary} In conclusion, the energy levels of alkali Rydberg states
emerging from the hydrogenic $n_0$-manifold clearly reflect the phase space
structure of the microwave driven Coulomb problem, despite the presence of a
symmetry breaking atomic core. Also the ionization rates of the atoms reflect
the underlying classical phase space structure, with the exception of local
enhancements due to multiphoton resonances with nonhydrogenic sublevels of
other manifolds. We have checked that 
the observed structure is robust under changes of the driving
field amplitude, up to values where adjacent $n$-manifolds start to overlap.

\stars
We thank Dominique Delande and Ken Taylor for fruitful discussions and an
introduction to the R-matrix approach of \cite{halley93}.

\end{document}

%% file: euromacr.tex
\def\And{{\rm and\ }}

\def\stars{\bigskip\centerline{***}\medskip}

\newif\ifboo \boofalse

\def\Review#1{\boofalse{\it #1},}
\def\Name#1{{\sc #1},}
\def\Vol#1{\ifboo Vol. {\bf #1}\else{\bf #1}\fi}
\def\Year#1{\ifboo #1\else(#1)\fi}
\def\Book#1{\bootrue{\it #1},}
\def\Page#1{\ifboo {\rm p. #1}\else{\rm #1}\fi}